\documentclass[prl,twocolumn,showpacs,superscriptaddress]{revtex4}
\usepackage{amsmath,amsfonts,paralist}
\usepackage{graphicx}
\usepackage{epstopdf}

\newcommand{\be}{\begin{equation}}
\newcommand{\ee}{\end{equation}}
\newcommand{\bq}{\begin{eqnarray}}
\newcommand{\eq}{\end{eqnarray}}

\newcommand{\bc}{\begin{center}}
\newcommand{\ec}{\end{center}}

\def\(({\left(}
\def\)){\right)}
\def\[[{\left[}
\def\]]{\right]}
\def\bi{\bibitem}
\def \lan{\langle}
\def\ran{\rangle}

\begin{document}

\title{On the scenario for the glass transition }
\author{Giorgio Parisi} 
\affiliation{Dipartimento di Fisica, Sapienza Universit\`a di Roma, INFN, Sezione di Roma I,
  Statistical Mechanics and Complexity Center (SMC) - INFM
  - CNR, P.le Aldo Moro 2, I-00185 Roma, Italy}

\begin{abstract}
In this letter we study a lattice glass system that undergoes a glass transition. When we approach the glass transition we find both a divergence of a point to set correlation length and a vanishing of the thermodynamic potential. These findings are similar to the predictions coming from mean-field (replica) theory, but they differ from these predictions in some details: they underline the need of a better theoretical understanding of the glass transition.

\end{abstract}


\maketitle
Second order phase transitions are usually characterized by the existence of  an equilibrium correlation length that is divergent at the phase transition point.
It is usually stated that such a correlation length is absent for the glass transition. 
However it was noticed long time that in the framework of the Adams Gibbs theory \cite{AdGibbs, KWT,CAV} the slow dynamics is dominated by large regions that move in a cooperative way.   This observation has been formalized by introducing the appropriate correlations functions that can be used to define the dynamical correlation length \cite{KT,DAS,FP,SH,PL}. A precise theoretical analysis has been done both in the framework the replica approach \cite{FP, srpsp,BB0,AM,XW} and in the framework of the mode-coupling theory \cite{BB}. 

It was later realized that we can define a static correlation length that is divergent at the transition temperature \cite{BB,AM,XW,LB,CGV}. It is related to the point to set correlations: its definition is the following. We start from a large system at equilibrium: we froze it and we allow only a region of size $R$ to move. We ask if the equilibrium configuration satisfying this constraint is very far from the original configuration. It is natural to suppose that if $R<<\xi_d$, where $\xi_d$ is the typical size of the dynamically rearranging regions,
the system is blocked. The opposite behaviour should be present for $R>>\xi_d$. We can define an equilibrium correlation length $\xi_s$ as the minimal size of a non blocked region: the previous arguments suggest that we should have $\xi_s\propto \xi_d$ . It has been shown \cite{AM} that in the mean field approach $\xi_s$ diverges near the transition point.
While real experiments to measure $\xi_s$ are not yet feasible, numerical experiments have been done showing evidence of a large increase of the $\xi_s$ near the thermodynamic transition point \cite{CGV}.

In this letter I will study a different correlation length $\xi_w$. The definition is similar  to $\xi_s$, but it is technically different \cite{KOB}: we froze the system on a wall (a two dimensional slice) and we measure how the final configuration is different from the initial one (the geometry is similar to the one used in \cite{KP} for the dynamics). More precisely we introduce an overlap $q$ that indicates the similarity of two configurations: $q=1$ for identical configuration while $q$ takes a small value for unrelated configuration. In this way, if we froze the region where $0<x<1$, $x$ being one of the coordinates, we can define the average value of $q$ at given $x$, i.e. $q(x)$. The wall correlation length is defined by
\be
q(x) \approx a \exp(-x/\xi_w)+q_{bulk}
\ee
at large $x$. The correlation length $\xi_w$ is logically different from $\xi_s$ and its behaviour in mean field theory is not clear.

In this letter we present evidence that the correlation length  $\xi_w$ is divergent when we approach the phase transition in a lattice glass model (a similar conclusion has been reached for an off lattice fluid \cite{KOB}). We will relate the divergence of $\xi_w$ to the development of flat direction for an appropriate thermodynamic potential \cite{REP}.

\begin{figure}[t!]
\includegraphics[width=1.0\columnwidth]{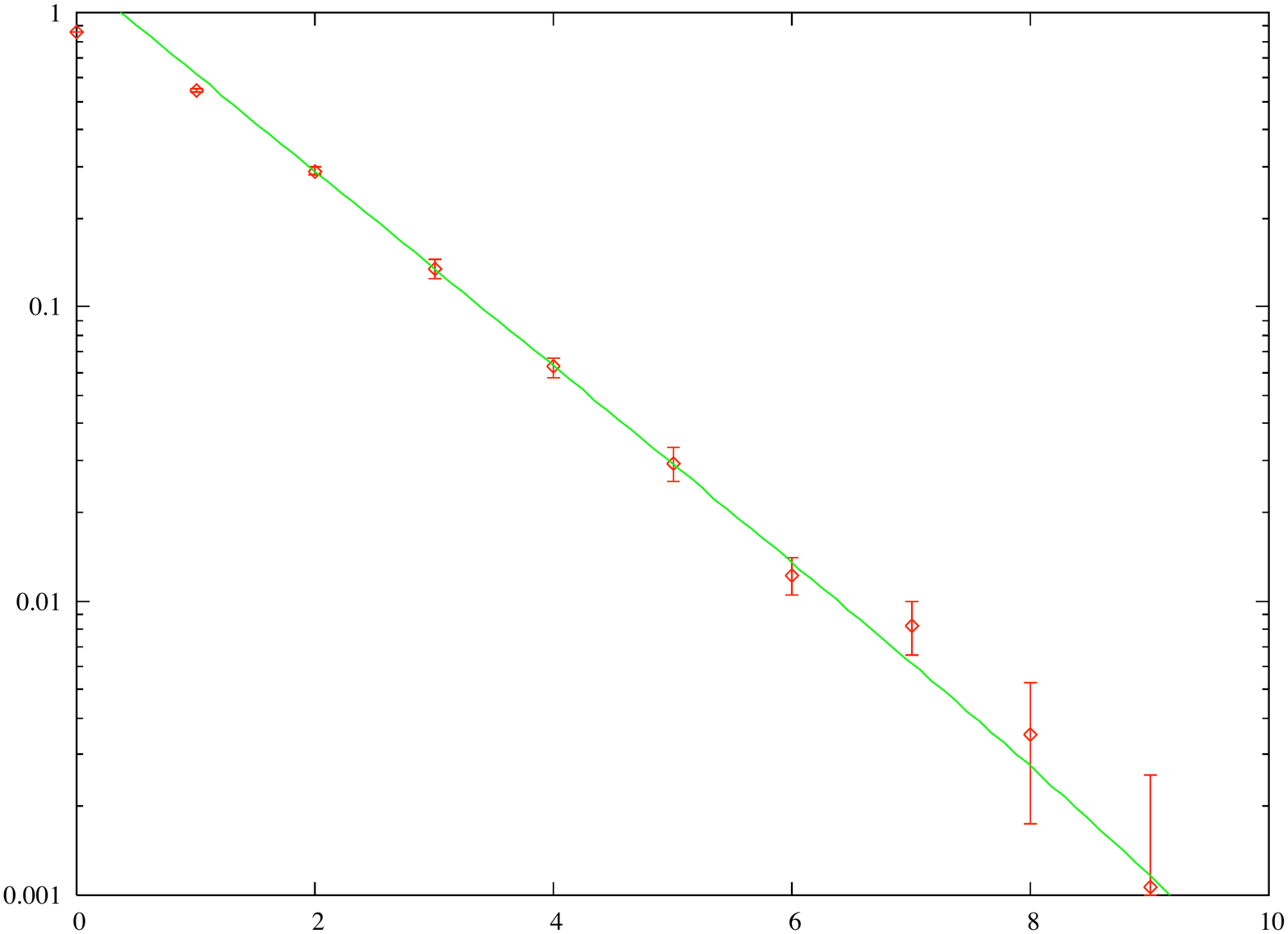}
\caption{The data for $q(x)-q_{bulk}$  at $\mu=8.5$ ($\rho=0.8097$)  as function of $x$ on a cubic lattice of side $L=22$. The value of the correlation length is $\xi_w=1.32 \pm 0.03$.}
\label{FIT}
\end{figure}
The model we consider is a simple lattice gas model (the CTCC model \cite{CTCC}),  that does not easily crystallize in simulations (the cell crystal contains $7^3=343$ particles and it is quite difficult to form). Each site  of a three dimensional cubic lattice may be empty or occupied. If the site is occupied we have to specify the orientation of the particle. The particles may point in each of the six directions that correspond to elementary moves on the lattice. The system satisfies the constraints: 1) each lattice point may occupied by only one kind of particles, 2) if the site $\vec{i}$ is occupied by a particle that points in the direction $\vec{d}$, the site $\vec{i}+\vec{d}$ is empty.

We introduce a chemical potential $\mu$. In the limit $\mu \to \infty$ we reach the maximum density state, i.e. a crystal with periodicity 7 and density 6/7=0.8571. In simulations on large lattices, starting from the disorder phase, there is no sign of crystallization up to the longest runs. One observes in the dynamics a very nice slowing down, with all the standard stigmata of a glass transition, i.e. mode-coupling behaviour of the correlation function (with both the $\alpha$ and the $\beta$ region well exposed), a strong increase of the dynamical susceptibilities and a simultaneous decrease of the the diffusion constant $D $ \cite{CTCC}.
This data on the diffusion constant do not have a particular nice behaviour as function of the chemical potential $\mu$. They are much more regular as function of the density. 

We will show that there are statically defined quantities that display a critical behaviour: they should be  counterparts of these dynamical quantities.
Let as start with a precise definition of $\xi_w$. We consider a large lattice of size $L$: the value of $L$ will not enter in the analysis but  consistency implies that $L>>\xi_w$. At given chemical potential $\mu$ we take a thermalized configuration $\sigma$. We now consider a system where at $x=0$  the configurations $\tau$ satisfy the constraint
 \be
\tau(0,y,z)=\sigma(0,y,z)\ .\label{CON}
\ee
In other words the particles on the wall (i.e. at $x=0$) cannot move and the particles outside the wall cannot penetrate inside.

We define 
\bq
q_{\sigma,\tau}(x,y,z)=\delta_{\sigma(x,y,z),\tau(x,y,z)}\, , \nonumber
\\ q_{\sigma}(x,y,z)= \lan q_{\sigma,\tau}(x,y,z)\rangle_\tau , \   q(x,y,z)= \lan q_{\sigma}(x,y,z)\rangle_\sigma .
\eq
Here $\lan\cdot\ran_\sigma$ and $\lan\cdot\ran_\tau$ denote respectively the statistical average over the variables $\sigma$ and $\tau$. We perform a double statistical average, first on the $\tau$ and later on the $\sigma$. Translational invariance implies that $ q(x,y,z)$ does not depend on $y$ and $z$ and it will be denoted by $q(x)$. This construction is similar to the one used in \cite{BB,AM,XW,LB,CGV} for defining $\xi_s$. Here the constraint eq. (\ref{CON}) is imposed on a slice an most of the system is free; in the definition of $\xi_s$ the same constraint is imposed on the whole system and only a small region is left free.
 
The function $q(x)$ tell us how much two equilibrium configurations are different if they coincide at distance $x$ in one direction. Which should be the behaviour of $q(x)$ near the thermodynamic glass transition? In the mosaic picture the function $q(x)$ should be strongly different from its asymptotic value at large $x$ in a region that becomes larger and larger when we approach the glass transition. The shape of the function $q(x)$ near the wall is an interesting problem that we do not address here. We focalize our interest on the behaviour at large $x$. We find that for $x\ge 2$ the function $q(x)$ can be remarkably well fitted by an exponential, i.e. 
\be
q(x)-q_{bulk}=a \exp(-x/\xi_w) +c
\ee
 where $c$ is a small constant, i.e. $O(10^{-3})$ (the previous hand-waving argument suggest that $\xi_w$ should diverges at the thermodynamic phase transition were $\xi_s$ diverges).
 An example of the fit is shown in fig.(\ref{FIT}), where the data are taken at $\mu=8.5$ ($\rho=0.810$) on a cubic lattice of side $L=22$. The value of the correlation length is $1.32 \pm 0.03$.

\begin{figure}[t!]
\includegraphics[width=1.0\columnwidth]{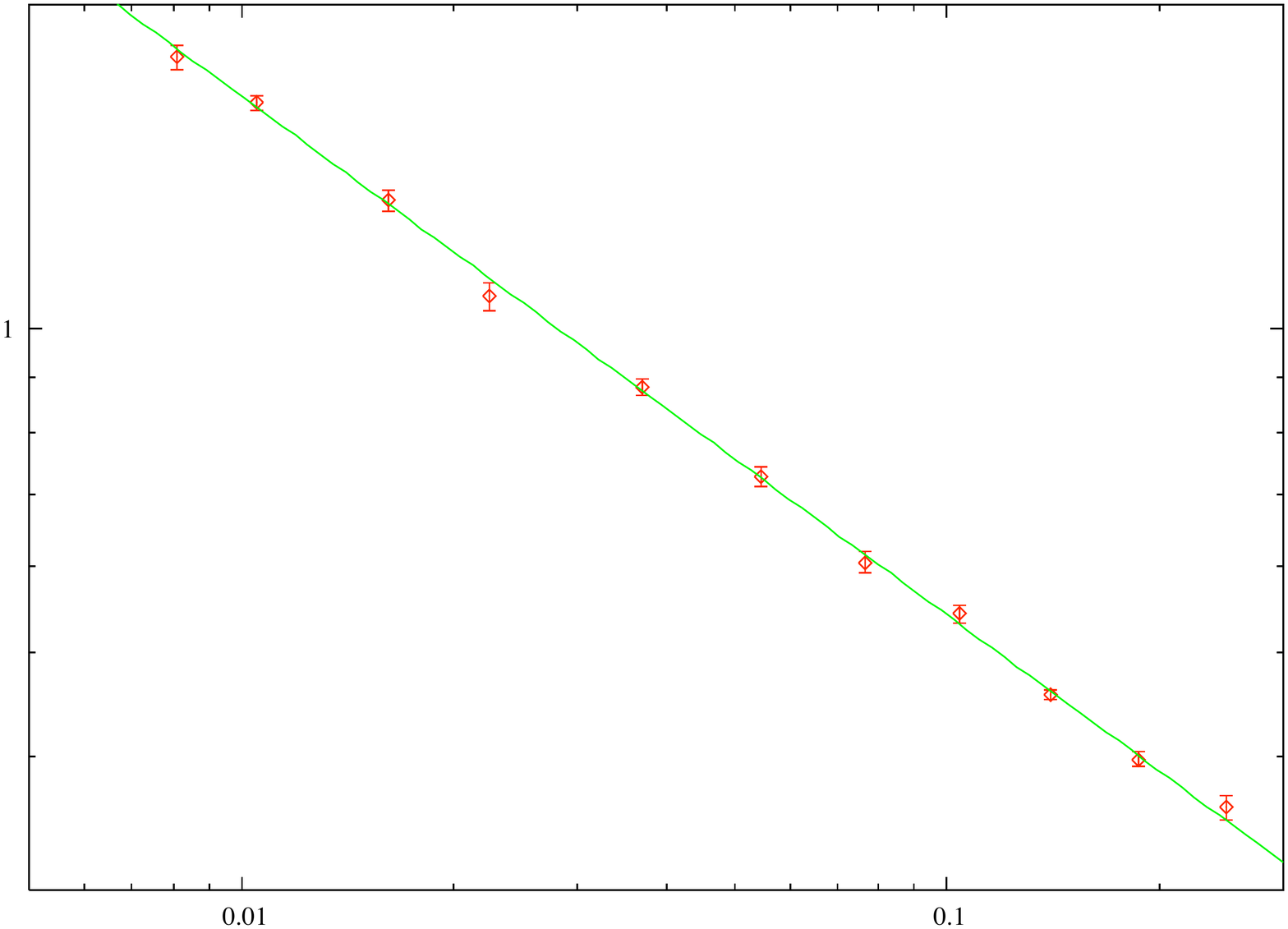}
\caption{The correlation length $\xi_w$ in the region of $\mu$ ranging from $1$ to $9.25$ as function of $\rho-\rho_c$, with $\rho_c=0.826$.}
\label{XI}
\end{figure}

We have measured this correlation (or penetration) length $\xi_w$ in the region of $\mu$ ranging from $1$ to $9.25$ on lattices of various side (up to $L=28$). The correlation $\xi_w$ is very well fitted by a power low behaviour
\be
\xi_w(\rho)=A(\rho_c-\rho)^{\nu}\ ,
\ee
with $\rho_c=0.826\pm 0.002$ and $\nu=0.48\pm0.02$ (see fig. (\ref{XI})). The value of the exponent $\nu$ is very near to the {\sl simple} value 1/2. 

\begin{figure}[t!]
\includegraphics[width=1.0\columnwidth]{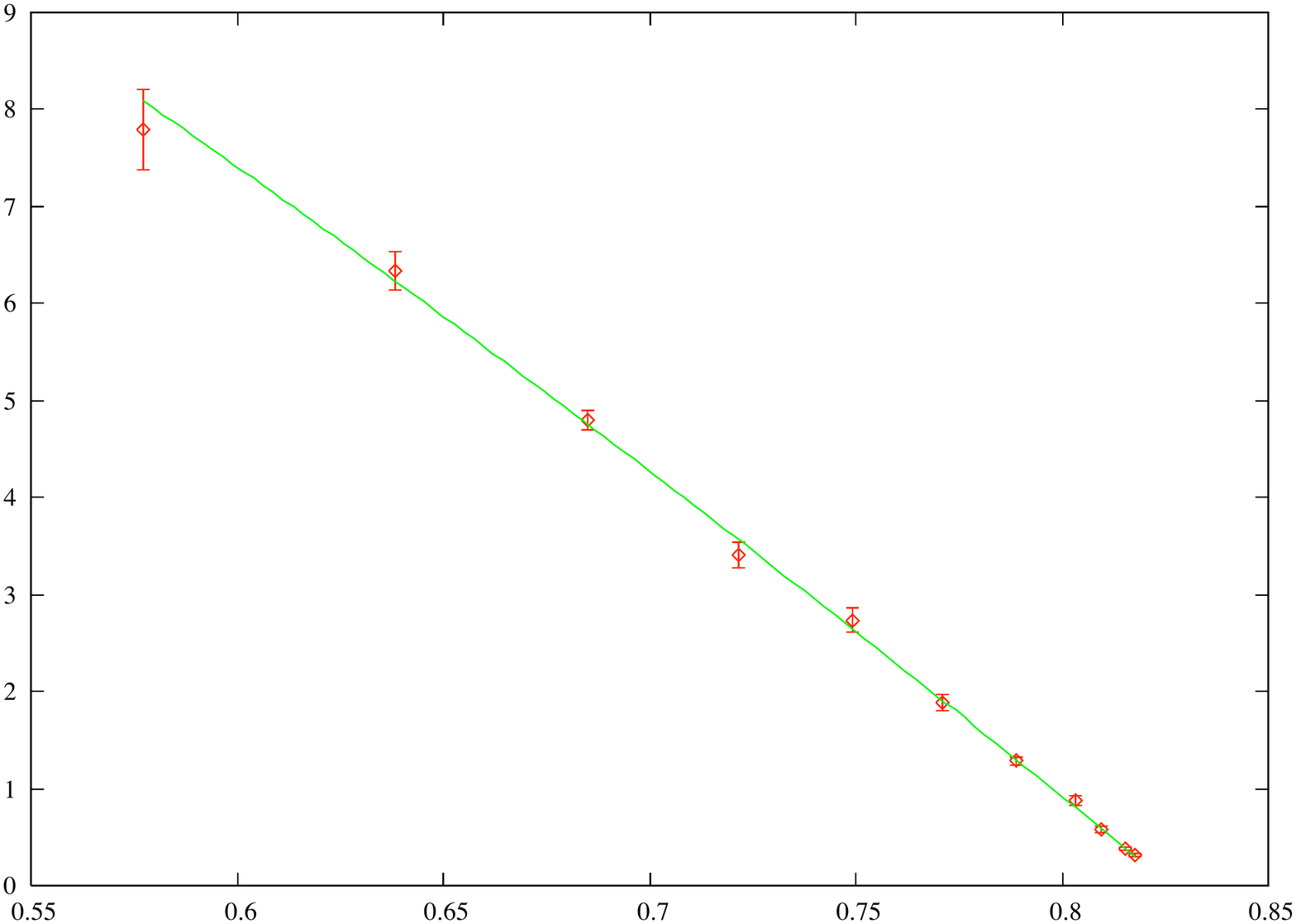}
\caption{ The same data as fig (\ref{XI}): here we plot $\xi_w(\rho)^{-2}$ versus $\rho$ in a linear scale.}
\label{LINEAR}
\end{figure}

In the study of phase transitions one usually introduces an order parameter and an associated thermodynamic potential. If the order parameter is a {\sl continuos}  function of the temperature, at  the phase transition point the potential is zero only in one point, while at lower temperatures it is different from zero in a region whose side goes to zero at the critical point. 
For a transition with a {\sl discontinuous} order parameter, at the phase transition point the potential is zero when the order parameter stays in a given interval and a similar behaviour is present also below the phase transition point. In the infinite volume limit the potential must be  convex function: when the potential is not convex in a mean field approximation, Maxwell construction enforces the convexity.

The glass transition may be characterized by the existence of regions of configuration space where (below the transition) the system remains trapped  for an infinitely large time. The overlap $q$ is the putative order parameter of the glass transition.  
 One can sharpen the physical picture by introducing a potential $W(q)$ defined as follows \cite{REP,MP,PZ}. We consider an equilibrium configuration $\sigma$.
We call $P_\sigma(q)$ the probability that an other configuration $\tau$ has an overlap $q_{\sigma,\tau}=q$. We define
\be
W(q)=-\lim_{V\to\infty}{\ln(P_\sigma(q))\over V} \ .
\ee
With probability one when the volume $V$ goes to infinity, the potential $W(q)$ does not depend on the reference configuration $\sigma$. In  other words $P_\sigma(q)\approx \exp (- V W(q))$.  By construction $W(q_{bulk})=0$ and the vanishing of the potential $W(q)$ for more than one $q$-value  is the distintive characteristic of replica symmetry breaking (it should happen below the thermodynamic glass transition).

\begin{figure}[t!]
\includegraphics[width=1.0\columnwidth]{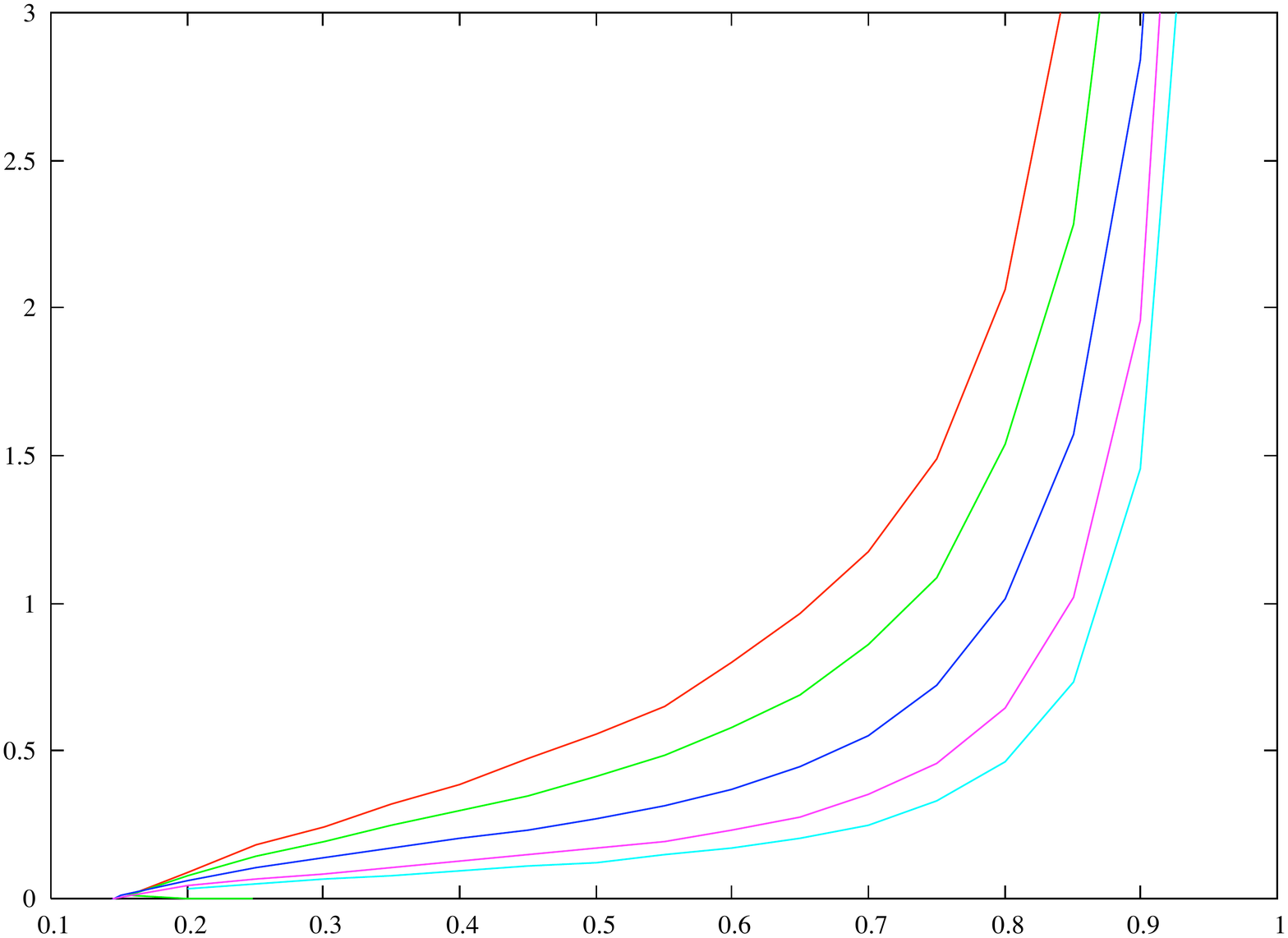}
\caption{The derivative of the potential $W'_\rho(q)$ at various densities (0. 749, 0.771,  0.789, 0.803,
0.810) on large systems of side $L=40$.}
\label{POT}
\end{figure}

There are many possible techniques that can be used to compute the potential $W(q)$. One of the best is the introduction of a tethering potential, as described in
\cite{V}. Here we use a simpler method \cite{NOI}: we constrain the $\tau$ configuration to stay in the region $q_{\sigma,\tau}>q^*$. With this constrain the resulting probability distribution of $q_{\sigma,\tau}$  is proportional to $\theta(q-q^*)\exp( -V W(q-q^*))$. In the region where $q^*>q_{bulk}$, we have that
\be
V \lan q_{\sigma,\tau}- q^* \ran= \left({dW\over dq}\right)^{-1}=1/W'(q^*)
\ee
Using a wild metaphor we could say that $W'(q)$ correspond to the force that pushes the system away from the region where $q$ is high. It is interesting to note that in the Maxwell construction $W'(q)$ is constant in the region where the naive mean field prediction violates convexity.

In the replica approach to glasses the potential $W(q)$ plays a central role \cite{FP}: in the mean field approach at high temperature the potential is convex. By decreasing the temperature firstly the convexity is lost; at a lower temperature (i.e. at the dynamic or mode-coupling transition) a secondary local minimum  ($q_m$) appears. The height of the minimum (i.e $W(q_{m})$) is the configurational complexity $\Sigma$. By decreasing the temperature the value of $W(q)$ at the secondary minimum decrease up to the point  (the Kauzmann transition)  where $W(q_m)\equiv \Sigma=0$. Below this temperature we are in the glass phase: two minima (with $W(q)=0$) are present and the replica symmetry is broken. This scenario is valid in infinite range model where very detailed computation can be done.

It is clear that the previous picture must be strongly modified in finite dimensions. We have already remarked the any thermodynamic potential in the infinite volume limit must be convex.  A consistent scenario could be provided by the Maxwell construction,
 unfortunately there are not many  numerical result on the behaviour of the potential in finite dimensional models \cite{FP}.

\begin{figure}[t!]
\includegraphics[width=1.0\columnwidth]{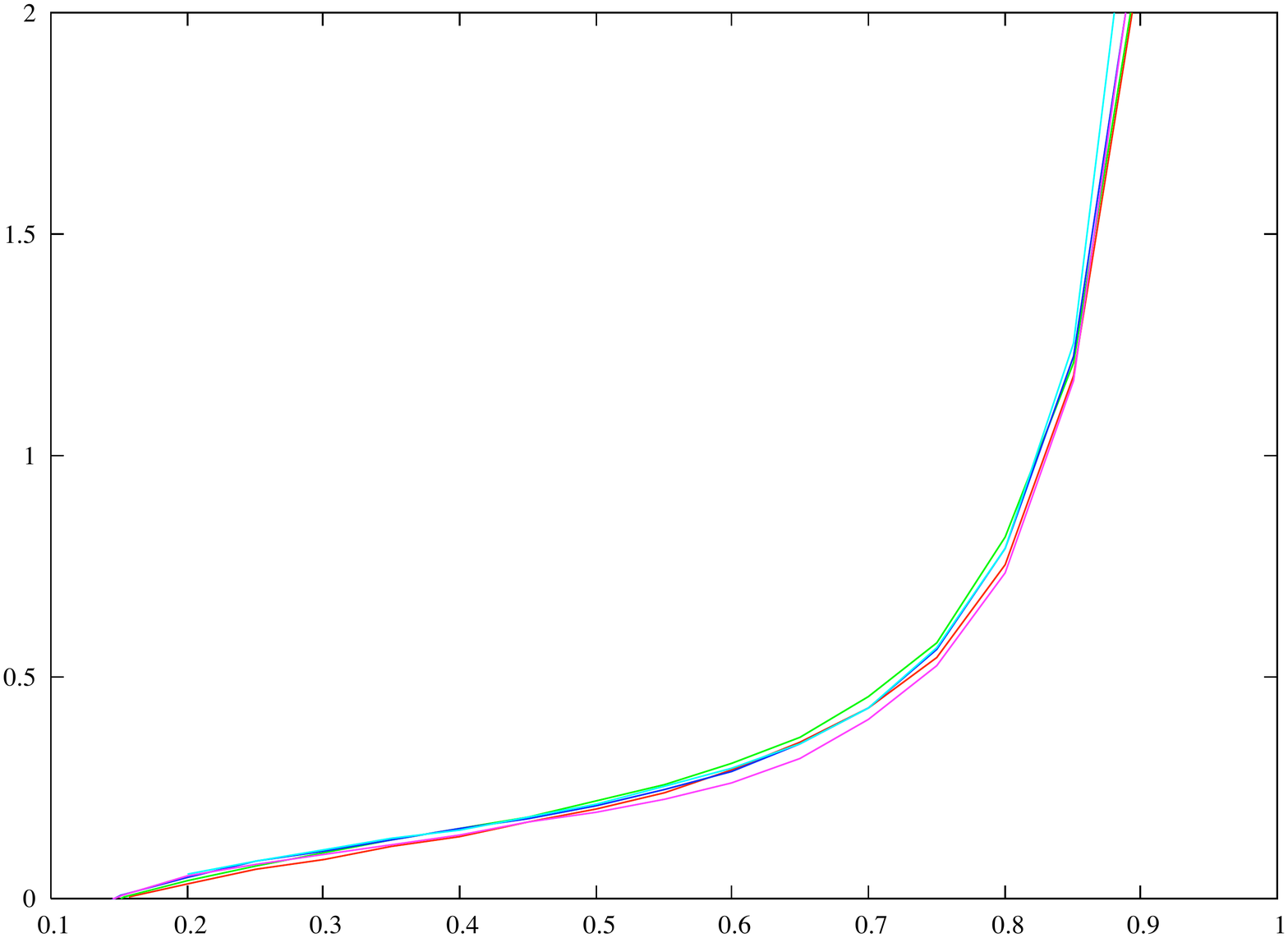}
\caption{The same data of fig. (\ref{POT}) for derivative of the potential $W'_\rho(q)$ multiplied by $\xi_w(\rho)^2$, i.e. $\Omega_\rho'(q)$.}
\label{POTS}
\end{figure}

In this letter we compute the the potential $W_\rho(q)$ in the CTCC lattice gas model and we find that it has some surprising properties. The derivative of the potential $W'(q)$ at various densities (from 0.749 to 0.810) is shown in figs. (\ref{POT}) and (\ref{POTS}). A few remarks are in order:
\begin{itemize}
\item We cannot identify regions  where $W'_\rho(q)$ decreases by increasing $q$: convexity is always strictly satisfied.
\item There are no flat regions  ($W'_\rho(q)= \mbox{const})$ that could be the remnant of the Maxwell construction. Therefore no dynamical (mode coupling) transition can be identified.
\item The potentials at different densities have similar shape.
\item The potential strongly decrease in nearly the whole $q$ interval when the density increases.
\end{itemize}

The last two statements can be done more sharply if we rescale the potential with the square of the correlation length 
\be
\Omega_\rho'(q)\equiv \xi_w(\rho)^2 W'_\rho(q) \ .
\ee
The data for $\Omega_\rho'(q)$ are shown in fig. (\ref{POTS}) and they are weakly dependent on the density. If  $\xi_w(\rho)$ goes to infinity at a critical value of $\rho$ the potential $W(q)$ goes to zero in the whole interval from $q_m$ up to a value of $q$ of near to $0.9$.

The results are surprising (at least to me) for the following reasons:
\begin{itemize}
\item The potential $W(q)$ cannot be used to identify the (dynamic) mode coupling transition: the configurational complexity cannot be defined. However it is possible that in this model the dynamic transition is very near to the static one, so that the two transitions cannot be easily disentangled: is this case one may observing overlapping features. 
One should also remember that $W(1)$ is the entropy. If  at the Kauzmann transition the entropy takes a very small value, the potential $W'(q)$ should be nearly vanishing in large  interval of $q$.
\item The extrapolation of these results implies that a  static Kauzmann transition is present and the behaviour is rather simple. The exponent $\nu$ is very near to the {\sl  simple} value $1/2$ and the thermodynamic $W_\rho(q)$ potential scales naively without any need of introducing extra power corrections.
\end{itemize}

Similar investigations are presently done also for off-lattice fluids \cite{KOB, NOI} and it would be very interesting to compare the results. A very important question is the theoretical derivation of these results.

I am happy to thank Andrea Cavagna,  Giulio Biroli, Chiara Cammarota, Silvio Franz, Irene Giardina, Giacomo Gradenigo, Thomas Grigera, Walter Kob, Paolo Verrocchio and Francesco Zamponi for very useful discussions.


\begin{thebibliography}{99}
\bibitem{AdGibbs} G. Adams and J.H. Gibbs J.Chem.Phys {\bf 43} (1965) 139.


\bi{KWT} T.R. Kirkpatrick and D. Thirumalai, Phys. Rev. Lett. {\bf 58},
2091 (1987); T R. Kirkpatrick and D. Thirumalai, Phys. Rev. {\bf B36}, 5388 (1987);
T.  R.  Kirkpatrick, D.  Thirumalai and P.G.  Wolynes, Phys.  Rev.  {\bf A40}, 1045 (1989).

\bi{CAV} A. Cavagna, Physics Reports {\bf 51} 476 (2009).

\bi{KT} T. R. Kirkpatrick and D. Thirumalai  Phys. Rev. A {\bf 37}, 4439 (1988)
\bi{DAS} C. Dasgupta, A. Indrani, S. Ramaswamy, and M. Phani,
Europhys. Lett.  {\bf 15} 307 (1991).

\bi{FP} S. Franz and G. Parisi, J. Phys. Cond. Mat.  {\bf 12}, 6335
(2000).
\bibitem{srpsp}
S. Franz and G. Parisi Euro.  Phys.  J. {\bf B 8} 417 (1999);  M. Campellone, G. Parisi and P. Ranieri, Phys.  
Rev.  {\bf B 59}.  1036 (1999).

\bibitem{SH}C. Donati, J.F. Douglas, W. Kob, S.J. Plimpton, P.H.
Poole and S.C. Glotzer, Phys. Rev. Lett.  {\bf 80}, 2338 (1998).
\bibitem{PL} G. Parisi  J. Phys. Chem. B. {\bf 103}, 4128 (1999).

\bibitem{BB0} J. Bouchaud and G. Biroli, J. Chem. Phys. {\bf 121}, 7347
(2004).

\bibitem{AM} A. Montanari and G. Semerjian, Phys. Rev. Lett. {\bf 94} 247201 (2005),
 J. Stat. Phys. {\bf 124} 103 (2006).

\bibitem{XW}
X. Xia and P. G. Wolynes,  Proc. Nat. Acad. Sci. {\bf 9}
2990 (2001); Phys. Rev. Lett. {\bf 86} 5526 (2001).

\bibitem{BB} J. Bouchaud and G. Biroli, Europhys. Lett. Vol 67  {\bf 21} (2004).

\bibitem{LB} V. Lubchenko and P. G. Wolynes, Ann. Rev. Phys.
Chem. {\bf  58}, 235 (2007).

\bibitem{CGV} A. Cavagna, T. S. Grigera, and P. Verrocchio, Phys.
Revi. Lett.  {\bf 98}, 187801 (2007); G. Biroli, J.-P. Bouchaud, A. Cavagna, T. S. Grigera,
P. Verrocchio, Nature Phys. {\bf 4} 771 (2008); C. Cammarota, A. Cavagna, G. Gradenigo, T. S. Grigera,
and P. Verrocchio,  arXiv:0904.1522 (2009) and arXiv:0906.3868 (2009).


\bibitem{KOB} S. Roldan-Vargas, L. Berthier, and W.Kob, (in preparation).

\bibitem{KP}
P. Scheidler, W. Kob, K.  Binder and G. Parisi, Phil. Mag. B {\bf 82}, 283 (2002).

\bibitem{REP} S. Franz, G. Parisi, J. Phys. I (France)  {\bf 5} (1995) 1401; Phys. Rev.
Lett. 79 (1997) 2486; Physica A  {\bf 261} (1998) 317.

\bibitem{CTCC} M. Pica Ciamarra, M. Tarzia, A. de Candia and A.
Coniglio, Phys. Rev. E {\bf 67} 057105 (2003); Phys. Rev.
E  {\bf 68}, 066111 (2003).
\bibitem{PZ} G. Parisi and F.Zamponi, {\em Mean field theory of hard sphere glasses and jamming},  arXiv:0802.2180, Rev. Mod. Phys. (in press).
\bibitem{MP} M. M\'ezard and G. Parisi, {\em Glasses and replicas},  arXiv:0910.2838.

\bibitem{BIN}  B. Coluzzi, M. M\'ezard,  G. Parisi and P. Verrocchio, J. Chem. Phys. {\bf 111} 9039 (1999).

\bibitem{AF} L. Angelani and G. Foffi,  J. Phys. Condens. Matter {\bf 19} 256207 (2007).

\bibitem{V} L.A. Fernandez, V. Martin-Mayor and D. Yllanes, Nucl. Phys. B {\bf 807} 424 (2009).


\bibitem{NOI}  A. Cavagna,  C. Cammarota, I. Giardina, G. Gradenigo, T. Grigera, G. Parisi  and P. Verrocchio, Poster presented at the $6^{th}$ IDMRCS, Rome, (2009) (unpublished) and in preparation.
\end{thebibliography}
\end{document}